\documentclass[ aps, pra, onecolumn, superscriptaddress, nofootinbib, longbibliography ]{revtex4-2}
\usepackage{amsmath}
\usepackage{amssymb}
\usepackage{graphicx}
\usepackage{xcolor}
\usepackage{hyperref}
\usepackage{tabularx}
\usepackage{multirow}

\usepackage{tikz}
\usetikzlibrary{positioning,arrows.meta}

\begin{document}
\title{Quantum Interconnects Part I: Strategic Quantum Network Formation}
\author{Gustavo Castro do Amaral}
\affiliation{The Netherlands Organization for Applied Scientific Research, Delft, The Netherlands}

\begin{abstract}
The realization of large-scale quantum networks requires more than advances in quantum repeaters, memories, and processors, it requires a framework explaining how heterogeneous quantum technologies evolve from isolated deployments into interconnected infrastructures. While the classical Internet evolved under strong utility incentives associated with resource sharing and communication demands, quantum networking currently lacks dominant applications capable of generating comparable incentives. As a consequence, contemporary quantum networks are largely formed through technology-driven decisions motivated by technical feasibility, experimental validation, and expected future value. This work argues that the absence of utility-driven network formation is not solely a consequence of immature applications, but also of insufficient abstraction. In particular, heterogeneous quantum platforms remain tightly coupled to the functionalities they provide, preventing the definition of technology-independent utility functions. A hierarchical architecture consisting of Physical Platforms (PP), Functionalities (F), Services (S), Applications (A), and Use-Cases (UC) is proposed, together with the argument that quantum interconnects constitute the enabling technology required to decouple physical implementations from network functionalities. Such decoupling permits the definition of utility functions at the functionality level and establishes the conditions under which strategic (agent-based) network formation becomes applicable. Quantum interconnects should therefore be viewed not only as interoperability devices, but also as fundamental enablers of strategic quantum network evolution.
\end{abstract}
\maketitle

\section{Introduction}

Entanglement is an operational resource capable of enabling applications whose performance exceeds that of their classical counterparts, including quantum communications, distributed quantum computing, and quantum sensing \cite{horodecki2009entanglement, wehner2018quantuminternet, cuomo2020distributed, degen2017quantumsensing}. Quantum networks arise from the need to distribute entanglement between geographically separated nodes and thereby enable such applications over distances exceeding those supported by direct transmission. A defining characteristic of quantum networking is that entanglement simultaneously represents both the resource consumed by the network and its primary objective. Entanglement swapping, for example, consumes locally established entanglement in order to create entanglement between distant parties \cite{zukowski1993entanglementswapping}. Consequently, a quantum network can be interpreted as an infrastructure that consumes short-range entanglement to establish long-range entanglement \cite{briegel1998repeaters, kimble2008quantuminternet, wehner2018quantuminternet}. 

Current worldwide efforts toward quantum networking exhibit a predominantly homogeneous character: most demonstrations and deployment roadmaps assume a common physical platform throughout the network \cite{kimble2008quantuminternet, wehner2018quantuminternet, pompili2021multinode}. While partially motivated by engineering simplicity, this trend also reflects the absence of mature quantum interconnects capable of enabling interoperability between heterogeneous platforms. The central observation motivating this work is that interoperability has consequences extending beyond network connectivity. Specifically, incompatibility between physical platforms prevents abstraction at the functionality level. Without such abstraction, utility functions become difficult to define, strategic decision-making becomes challenging to formalize, and network formation remains primarily technology-driven. This contrasts to historical evidence that large-scale infrastructures evolve through strategic decisions made by independent stakeholders. Therefore, investigating the conditions under which strategic network formation becomes applicable to quantum networking is paramount to its success.

The thesis of this work can be stated as follows: \begin{quote} Quantum interconnects do not merely connect heterogeneous physical platforms. They enable functionality-level abstraction, which in turn enables utility definition and strategic quantum network formation. \end{quote} The thesis is supported by three hypotheses:
\begin{description}
\item[H1.] \textbf{Functional Abstraction Hypothesis}: Network-relevant properties emerge primarily at the functionality level. Consequently, large-scale quantum networks require the decoupling of physical platforms from the functionalities they provide.
\item[H2.] \textbf{Interconnect-Enabled Abstraction Hypothesis}: Quantum interconnects constitute the enabling mechanism that allows physical platforms to be abstracted as functionalities by providing interoperability between heterogeneous technologies.
\item[H3.] \textbf{Agent-Based Formation Hypothesis}: Once functionalities become platform-independent and interoperable, quantum network formation becomes increasingly utility-driven and can therefore be described through agent-based models.
\end{description}
Together, these hypotheses establish a conceptual path from platform interoperability to strategic network formation and represents a shift from homogeneous to heterogeneous quantum network design.

\section{Network Formation}

Network formation has historically been studied from two distinct perspectives \cite{jackson2010social}. The first consists of dynamic growth models, in which networks evolve through the repeated addition of nodes and links \cite{erdos1960randomgraphs, watts1998collective, barabasi1999emergence}. A prominent example is the preferential attachment paradigm, where new nodes preferentially connect to already well-connected nodes. Such mechanisms naturally generate scale-free network structures and have been widely used to describe the growth of the classical Internet. The second consists of agent-based models, in which links emerge through strategic decisions made by autonomous actors \cite{jackson1996strategic, bala2000networkformation}. In this framework, network participants evaluate benefits and costs associated with connectivity and modify the network accordingly. The resulting structure emerges as the outcome of many decentralized decisions.

The early Internet possessed a clear incentive structure: organizations connected because doing so immediately increased access to information, computational resources, users, and content. As adoption increased, the utility associated with joining the network also increased, naturally reinforcing preferential attachment mechanisms. Quantum networks currently occupy a different position: most deployments are motivated by technological demonstration, scientific validation, consortium objectives, government programs, and expectations regarding future applications. To highlight the difficulty in designing a quantum network based on the use-case it could deliver, take the example of \textbf{Cybersecurity}, \textbf{Secure Elections}, and \textbf{Blind Dataset Classification}: although each can be associated with a quantum network application (Quantum Key Distribution, Quantum Position Verification, and Distributed Quantum Computing, respectively), theoretical and experimental support for the requirements and benefits are both at significantly different stages. This point is summarized and associated with a potential timeline in Table \ref{table:usecases}.

\begin{widetext}

\begin{center}
\begin{table}[h]
    \centering
    \caption{Use-Case Timeline Classification} \label{table:usecases}
    \setlength\tabcolsep{5pt}
    \footnotesize\centering
    \begin{tabular}{c|c|c|c}
    \multirow{2}{9em}{Use-Case\\Class} & \multirow{2}{9em}{Ready\\ ($x<5$ years)} & \multirow{2}{9em}{Medium-Term\\ ($5<x<10$ years)} & \multirow{2}{9em}{Long-Term\\ ($10<x$ years)}\\
    & & &\\
    \hline
    \hline
      \multirow{4}{9em}{Example} & \multirow{4}{9em}{Cybersecurity (QKD) \cite{gisin2002qkd}} & \multirow{4}{9em}{Secure Elections (QPV) \cite{buhrman2014qpv}} & \multirow{4}{9em}{Blind Dataset Classification\\ (Distributed QC) \cite{cuomo2020distributed}}\\
      & & &\\
      & & &\\
      & & &\\
      \multirow{2}{9em}{Benefits} & \multirow{2}{9em}{Known} & \multirow{2}{9em}{Not well-defined} & \multirow{2}{9em}{Not well-defined}\\
      & & &\\
      \multirow{2}{9em}{Requirements} & \multirow{2}{9em}{Known} & \multirow{2}{9em}{Known} & \multirow{2}{9em}{Not well-defined}\\
      & & &\\
    \hline
    \hline
    \end{tabular}
\end{table}
\end{center}
\end{widetext}

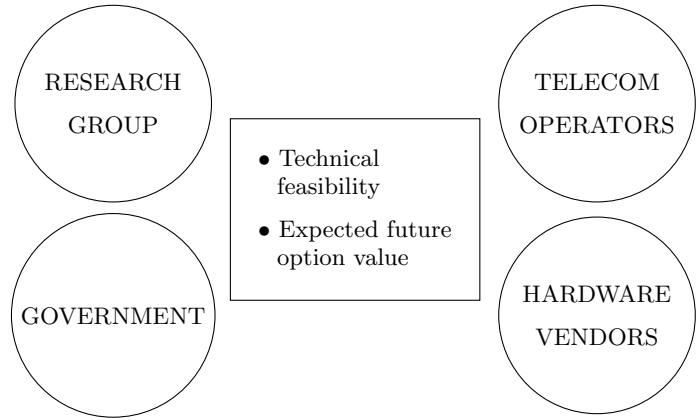
\begin{figure}[h]
\begin{tikzpicture}
[ stakeholder/.style={ draw, circle, minimum size=2.6cm, align=center }, mainbox/.style={ draw, rectangle, minimum width=3.3cm, minimum height=2.4cm, align=left } ]
\node[stakeholder] (lab) at (-3.2,1.4) {RESEARCH\\[2mm]GROUP}; \node[stakeholder] (telco) at (3.2,1.4) {TELECOM\\[2mm]OPERATORS}; \node[stakeholder] (gov) at (-3.2,-1.4) {GOVERNMENT}; \node[stakeholder] (vendor) at (3.2,-1.4) {HARDWARE\\[2mm]VENDORS};
\node[mainbox] (center) at (0,0) { $\bullet$ Technical\\ \hspace*{0.25cm}feasibility\\[2mm] $\bullet$ Expected future\\ \hspace*{0.25cm}option value };
\end{tikzpicture}
\label{fig:current}
\caption{Current strategical landscape of quantum network actors and their motivations.}
\end{figure}

An important outcome of the analysis of Table \ref{table:usecases} is: evaluating the impact of new technologies (physical platforms for quantum communications) based on the use-cases they could enable is largely inviable. Consequently, connectivity decisions are often justified through what may be called the \emph{``if this works''} argument. Consider four representative actors and the current motivations that drive quantum network formation. Each actor in Fig. 1 invests because future utility may emerge if quantum networking succeeds. In such a setting, network formation is primarily technology-driven rather than utility-driven. Ultimately, however, each stakeholder seeks to optimize specific objectives. Laboratories seek resource accessibility; telecom operators seek reachability; hardware vendors seek adoption; governments seek strategic capability and economical competitiveness. This observation motivates a transition from technology-driven deployment toward utility-driven deployment.

The canonical example of utility-driven network formation is the Jackson-Wolinsky (JW) connections model \cite{jackson1996strategic}. In this framework, utility may be expressed as
\begin{equation}
U_i(G) = \sum_{j \neq i} \delta^{d_{ij}} - c k_i ,
\end{equation}
where $d_{ij}$ denotes network distance, $k_i$ denotes the number of maintained links, $c$ represents link cost, and $\delta$ captures value decay with distance. The importance of this formulation is not its specific mathematical form, but rather the principle it embodies: networks form because autonomous actors perceive utility. Hence, the original picture shifts to the one depicted in Fig. 2:

\begin{figure}[h]
    \begin{tikzpicture}
    [ stakeholder/.style={ draw, rectangle, minimum width=2.8cm, minimum height=1.6cm, align=center } ]
    \node[stakeholder] (lab) {Hardware\\Vendors}; \node[stakeholder, right=1cm of lab] (gov) {Telecom\\Operators}; \node[stakeholder, right=1cm of gov] (telco) {Research\\Group}; \node[stakeholder, right=1cm of telco] (hw) {Government};
    \draw[->, thick] (hw.south) -- ++(0,-0.8); \draw[->, thick] (telco.south) -- ++(0,-0.8); \draw[->, thick] (lab.south) -- ++(0,-0.8); \draw[->, thick] (gov.south) -- ++(0,-0.8); 
    \node[align=center, below=1.0cm of lab] {Optimize\\quantum resources\\and accessibility};
    \node[align=center, below=1.0cm of gov] {Optimize strategic\\capability and economical\\competitiveness};
    \node[align=center, below=1.0cm of telco] {Optimize\\reachability}; 
    \node[align=center, below=1.0cm of hw] {Optimize hardware\\adoption}; 
    \end{tikzpicture}
    \label{fig:expected}
    \caption{Expected strategical landscape for quantum network actors and their motivations under a utility-driven\\network formation paradigm.}
\end{figure}
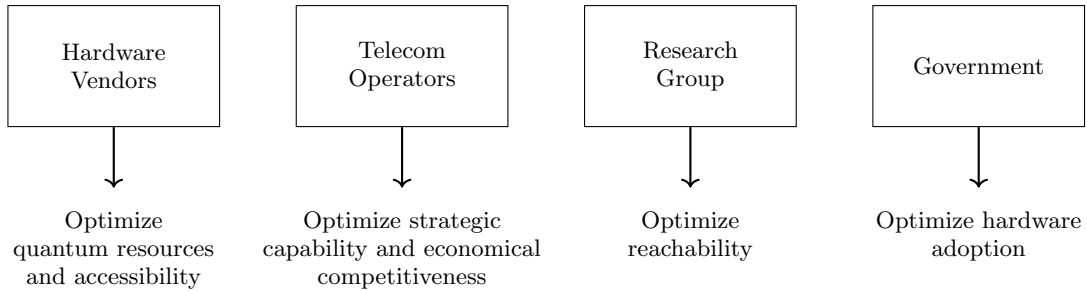

The challenge for quantum networking is therefore not the absence of a network formation framework. The challenge is the definition of utility itself. This raises the central question:
\begin{quote}
How does an agent assign utility to a quantum network when the underlying physical platforms are mutually incompatible?
\end{quote}

\section{Quantum Interconnects and Hierarchical Abstraction}

Consider a network composed of heterogeneous technologies, such as quantum dots, colour-centres, trapped ions, and neutral atoms \cite{kimble2008quantuminternet, wehner2018quantuminternet}. A utility function expressed directly in terms of physical platforms may take the form
\begin{equation}
U_i = f(\mathrm{Quantum Dots},\mathrm{Colour Centres},\mathrm{Trapped Ions},\mathrm{Cold Atoms},\ldots).
\end{equation}
Such formulations present a fundamental difficulty: every time a new platform appears, the utility function itself must be revised. Consequently, utility becomes dependent on implementation details rather than network capabilities. An alternative is to reason in terms of functionalities. We propose the hierarchy
\begin{equation}
 \boxed{ PP \rightarrow F \rightarrow S \rightarrow A \rightarrow UC },
\end{equation}
where PP: Physical Platforms, F: Functionalities, S: Services, A: Applications, UC: Use-Cases \cite{vanmeter2014quantumnetworking, dahlberg2019linklayer}. This abstraction is accompanied with definitions and examples, presented in Table \ref{table:hierarchy}; In Table \ref{table:examples}, we build the hierarchical chain in the reverse order, i.e., starting from a Use-Case and progressing all the way to the Platforms that enable it.

\begin{widetext}
\begin{center}
\begin{table}[h]
    \centering
    \caption{Hierarchical Class Definitions and Examples within the proposed classes} \label{table:hierarchy}
    \setlength\tabcolsep{5pt}
    \footnotesize\centering
    \begin{tabular}{c|c|c|c|c|c}
    \multirow{2}{5em}{Hierarchical\\Class} & \multirow{2}{14em}{Platforms} & \multirow{2}{8em}{Functionalities} & \multirow{2}{8em}{Services} & \multirow{2}{8em}{Applications} & \multirow{2}{8em}{Use-Cases}\\
    & & & &\\
    \hline
    \hline
      \multirow{8}{5em}{Definition} & \multirow{8}{14em}{\scriptsize{Physical systems capable of enabling quantum communications of the interfacing between quantum communication systems}} & \multirow{8}{8em}{\scriptsize{Fundamental operations enabled by the combination of platforms or by a single platform}} & \multirow{8}{8em}{\scriptsize{Economically-exploitable activities in the network}} & \multirow{8}{8em}{\scriptsize{Combination of services towards general operational resource of the network}} & \multirow{8}{8em}{\scriptsize{Activities made available to the user by a quantum network that generates value, either commercial or societal}}\\
      & & & &\\
      & & & &\\
      & & & &\\
      & & & &\\
      & & & &\\
      & & & &\\
      & & & &\\
      \multirow{9}{5em}{Examples} & \multirow{9}{14em}{\scriptsize{\textbf{Quantum Communication:} Absorptive Quantum Memories,\\ Emissive Quantum Memories, Entangled-Photon Sources\\ \textbf{Quantum Interconnects:} Quantum Frequency Converters, Spectral/Temporal Wave-Packet Shapers, Photonic Qubit Encoding Mappers}} & \multirow{9}{8em}{\scriptsize{Generation, Storage, Routing, Synchronization, Processing}} & \multirow{9}{8em}{\scriptsize{Storage Beyond Delay Line, Quantum-Memory-Assisted Entanglement Swapping, Concatenation of Elementary Links}} & \multirow{9}{8em}{\scriptsize{Blind Quantum Computing, Distributed Quantum Computing, Remote Quantum Sensing, Long-Distance Quantum Key Distribution}} & \multirow{9}{8em}{\scriptsize{Clock Synchronization, Cybersecurity, Secure Elections, Blind Dataset Classification}}\\
      & & & &\\
      & & & &\\
      & & & &\\
      & & & &\\
      & & & &\\
      & & & &\\
      & & & &\\
      & & & &\\
    \hline
    \hline
    \end{tabular}
\end{table}

\begin{table}[h]
    \centering
    \caption{Examples of hierarchical classes chained together} \label{table:examples}
    \setlength\tabcolsep{5pt}
    \footnotesize\centering
    \begin{tabular}{c|c|c}
    \multirow{2}{6em}{Hierarchy} & \multirow{2}{24em}{Example 1} & \multirow{2}{24em}{Example 2} \\
    & & \\
    \hline
    \hline
      \multirow{2}{6em}{Use-Case} & \multirow{2}{24em}{Blind Dataset Classification} & \multirow{2}{24em}{Cybersecurity} \\
      & & \\
      \multirow{2}{6em}{Application} & \multirow{2}{24em}{Blind Quantum Computing} & \multirow{2}{24em}{Long-Distance Quantum Key Distribution} \\
      & & \\
      \multirow{2}{6em}{Services} & \multirow{2}{24em}{Quantum-Memory-Assisted Entanglement Swapping, Storage Beyond Delay Line} & \multirow{2}{24em}{Storage Beyond Delay Line,\\Concatenation of Elementary Links} \\
      & & \\
      \multirow{2}{6em}{Functionality} & \multirow{2}{24em}{Entanglement Generation, Storage} & \multirow{2}{24em}{Storage, Routing} \\
      & & \\
      \multirow{5}{6em}{Platform} & \multirow{5}{24em}{\begin{itemize}
      \item Absorptive Quantum Memory
      \item Entangled-Pair Source
      \item Photonic Qubit Mapper
      \end{itemize}} & \multirow{5}{24em}{\begin{itemize}
      \item Emissive Quantum Memory
      \item Quantum Frequency Converter
      \end{itemize}} \\
      & & \\
      & & \\
      & & \\
      & & \\
    \hline
    \hline
    \end{tabular}
\end{table}
\end{center}
\end{widetext}

Utility can then be written as
\begin{equation}
U_i = \sum_j V_jF_{ij} - \sum_l C_l,
\end{equation}
where $V_j$ denotes functionality value, $F_{ij}$ represents achievable performance, and $C_l$ denotes deployment cost. Importantly, this formulation is independent of the underlying physical implementation. A technology may be replaced while preserving the functionality exposed to the network. This observation supports H1.

\subsection{Quantum Interconnects as Abstraction-Enabling Devices} 

Different quantum technologies exhibit distinct wavelengths, bandwidths, coherence times, interaction mechanisms, and encoding schemes. Consequently, functionality-level reasoning becomes difficult unless interoperability exists. Traditionally, interconnects are viewed as translation devices that enable communication between otherwise incompatible technologies \cite{askarani2021multiplatform}. We propose a broader interpretation. A quantum interconnect is an \emph{abstraction-enabling device}. Its significance is not merely that it connects platforms. Its significance is that it enables functionalities to become independent of physical realization. Conceptually,
\begin{equation}
PP \overset{\mathrm{Interconnect}}{\longrightarrow} F,
\end{equation}
an observation that supports H2.

The utility formulation can therefore be modified as
\begin{equation}
U_i = \sum_j V_j F_{ij} I_{ij} - \sum_l C_l,
\end{equation}
where $I_{ij} \in [0,1]$ represents interoperability. When two platforms cannot exchange resources, $I_{ij}=0$. When interoperability is perfect, $I_{ij}=1$. Interoperability thus appears naturally as a parameter influencing utility and network formation. This observation supports H3.

\section{From Technology-Driven to Agent-Based Formation}

The preceding arguments suggest that quantum network evolution may be understood through three developmental phases.
\begin{center}
\textbf{Phase 1: Technology-Driven Formation}
\end{center}
\vspace{-5pt}
\begin{center}
\begin{minipage}{0.8\textwidth}
Connectivity decisions are determined primarily by technical feasibility, funding availability, and experimental objectives. Utility is difficult to define and network growth is largely imposed rather than emergent.
\end{minipage}
\end{center}
\begin{center} \rule{0.8\textwidth}{0.4pt} \end{center}
\begin{center}
\textbf{Phase 2: Functionality-Abstraction Phase}
\end{center}
\vspace{-5pt}
\begin{center}
\begin{minipage}{0.8\textwidth}
Stakeholders increasingly describe network capabilities in terms of memory, generation, routing, and processing functionalities. However, interoperability remains limited and functionality abstraction exists primarily as an architectural concept.
\end{minipage}
\end{center}
\begin{center} \rule{0.8\textwidth}{0.4pt} \end{center}
\begin{center}
\textbf{Phase 3: Agent-Based Formation}
\end{center}
\vspace{-5pt}
\begin{center}
\begin{minipage}{0.8\textwidth}
Interoperable functionalities become available. Stakeholders can evaluate \begin{align*} \Delta U > \Delta C \end{align*} when considering new links, services, or deployments. Connectivity decisions become strategic and network growth increasingly reflects utility optimization. At this stage, agent-based network formation becomes an appropriate model for quantum networks.
\end{minipage}
\end{center}

\section{Conclusion}

This work argues that the principal obstacle preventing utility-driven quantum network formation is not merely the absence of mature applications, but the absence of sufficient abstraction. Three interconnected hypotheses were proposed. First, network-relevant properties emerge primarily at the functionality level rather than at the level of physical platforms. Second, quantum interconnects provide the interoperability necessary for functionality-level abstraction. Third, functionality-level abstraction enables the definition of utility functions and therefore allows quantum network formation to be modeled through agent-based mechanisms. These hypotheses establish the following causal chain (Fig. \ref{fig:hypotheses}): 

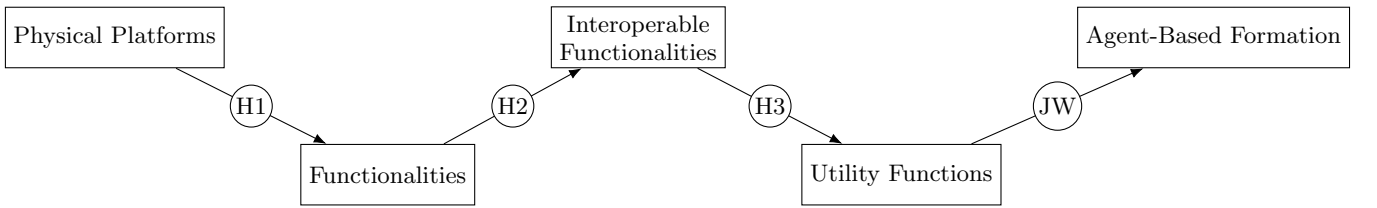
\begin{figure}[h]
\centering
\begin{tikzpicture}
[ node distance=2.8cm, box/.style={ draw, rectangle, align=center, minimum height=8mm }, >={Latex} ]
\node[box] (p) {Physical Platforms};
\node[box, below right=1cm and 1cm of p] (f) {Functionalities};
\node[box, above right=1cm and 1cm of f] (i) {Interoperable\\Functionalities};
\node[box, below right=1cm and 1cm of i] (u) {Utility Functions};
\node[box, above right=1cm and 1cm of u] (a) {Agent-Based Formation};
\draw[->] (p) -- node[midway,circle,draw,fill=white,inner sep=1pt] {H1} (f);
\draw[->] (f) -- node[midway,circle,draw,fill=white,inner sep=1pt] {H2} (i);
\draw[->] (i) -- node[midway,circle,draw,fill=white,inner sep=1pt] {H3} (u);
\draw[->] (u) -- node[midway,circle,draw,fill=white,inner sep=1pt] {JW} (a);
\end{tikzpicture}
\caption{Conceptual framework proposed in this work. H1 states that network-relevant properties emerge at the functionality level rather than the physical platform level. H2 states that quantum interconnects enable the abstraction of heterogeneous platforms into interoperable functionalities. H3 states that interoperable functionalities enable the definition of utility functions, allowing quantum network formation to transition from technology-driven deployment to agent-based formation under the JW model \cite{jackson1996strategic}.}
\label{fig:hypotheses}
\end{figure}

Under this interpretation, quantum interconnects are more than communication devices. They are foundational technologies that transform heterogeneous physical implementations into strategic network resources. By enabling the decoupling of physical platforms from functionalities, they create the conditions under which quantum networks can evolve from technology-driven demonstrations into self-organizing infrastructures governed by utility, incentives, and strategic decision-making. Finally, it is important to note that a network architecture and a network formation model are different. The proposed hierarchy does not by itself determine how networks form; rather, it provides the abstraction required for utility functions to be defined. Strategic quantum network formation then emerges from the decisions of stakeholders operating on those utilities.

\bibliographystyle{apsrev4-2} 
\bibliography{references}

\end{document}